# Remote Monitoring of Patient Respiration with Mask Attachment - A Pragmatic Solution for Medical Facilities

Vivian C. A. Koh[1,*], Yi Yang Ang[1], Wee Ser[2], and Rex X. Tan[1,2]

[1]R&D Department, Aevice Health Pte Ltd, Singapore
[2]School of Electrical and Electronic Engineering, Nanyang Technological University, Singapore
[*]vivian@aevice.com

**Abstract**

Remote monitoring of vital signs in infectious patients minimizes the risks of viral transmissions to healthcare professionals. Evidence indicates that donning face masks reduces the risk of viral transmissions and is now the norm in medical facilities. We propose attaching an acoustic-sensing device onto face masks to assist medical facilities in monitoring patients' respiration remotely. Usability and functionality studies of the modified face mask were evaluated on 16 healthy participants, who were blindfolded throughout the data collection. Around half of the participants noticed the difference between the modified and unmodified masks but they also reported there was no discomfort in using the modified mask. Respiratory rates of the participants were evaluated for one minute and the mean error of respiratory rate was found to be $2.0 \pm 1.3$ breath per minute. As all participants were healthy, the wheeze detection algorithm was assessed by playing 176 wheezes and 176 normal breaths through a foam mannequin. The recordings were played at three different times to account for varying environmental noise. The overall accuracy of the wheeze detection algorithm was 91.9%. The current findings support and suggest the use of the mask attachment in medical facilities.

## Introduction

At the time of writing this article, the world is combating a pandemic, Coronavirus Disease 2019 (COVID-19). New medical sub-industries like telehealth have been widely and rapidly adopted as mainstream solutions[1,2]. Many technologies have extended the applications' original intentions and have contributed positively to elevate the current situation. The health authorities now emphasize crowd management, patient monitoring, and revamping workflow to effectively utilize clinical resources and be prepared for any future events that can again overwhelm the system. Naturally, technology is one of the key enablers. Many technologically advanced facilities have equipped themselves with the latest medical devices for vital sign monitoring; some even started adopting wearable devices and mobile applications for patient tracking and remote clinics[3].

Donning face masks has also become a new norm, both in and out of healthcare facilities. The World Health Organization (WHO) recommends individuals to wear a face mask in all public places as a mid-term measure in combating the spread of virus[4]. Since, face masks of different varieties came under the spotlight of researchers, businesses, and consumers alike, scrutinizing their effectiveness in protecting the wearer from the virus or preventing the wearer from becoming a source of infection if unfortunately, infected[5,6].

Nevertheless, recent evidence suggests that face masks are integral in deterring the spread of COVID-19[7–9] when high compliance amongst the population is achieved[3,10,11]. In this context, Pan et al. proposed embedding a series of electronic sensors into a face mask to monitor blood oxygen saturation, heart rate, and temperature[12]. These vital physiological parameters are helpful to clinicians, especially when dealing with patients suffering from respiratory diseases. In line with these efforts, this article proposed a sensor device suitable for attaching on face masks, which listens to the wearer's breath, hence continuously monitors respiratory rate (RR) and detects any occurrence of wheeze (WZ). A detection in abnormal RR range (i.e. the normal range being 12-20 breaths per minute (brpm)) and WZ occurrence can indicate respiratory diseases such as COVID-19 pneumonia, asthma, and COPD exacerbation.

This work extends the recently proposed chest-worn wearable that addresses chronic respiratory disease care[13] and illustrates the applicability in repurposing the system as a mass patient management and respiratory monitoring aid. The chest-worn wearable device was modified and repurposed for this work and is known in this article as the "mask attachment". To the furthest extent of our knowledge, this is the first report of integrating an acoustic sensing mask attachment to medical face masks for the purpose of mass patient monitoring.

# Results and Discussion

## Usability

The success of a human-centered product design for any functional mask attachment can loosely be represented by the inability of the wearer to differentiate between wearing a mask with and without the attachment or experience any hindrance in human performance. As a qualitative analysis, 16 participants were asked to describe their experience according to the study design described in the latter section, which involved deprived visual-sensory interaction of the participants with two masks, one of which was modified with the mask attachment.

Ten out of sixteen participants felt that the two masks were different. One participant expressed that the "motion of the sensor" was felt for the modified mask and another reported that the sensor could be felt on the unmodified mask. As the mask attachment is a non-moving part and that the unmodified mask does not contain the mask attachment, these two participants' experiences were inconsistent with the facts. Three participants reported the unmodified mask was more breathable while two others reported the opposite. One of the two who reported the opposite said it may be due to the "tightness of the mask" worn on the unmodified mask. However, the tightness was not mentioned prior to the data collection session.

Five participants correctly reported the modified mask was slightly heavier than the unmodified mask. The actual weights of the unmodified mask and modified mask were 8 g and 14 g, respectively. Nevertheless, no participant reported any significant discomfort experienced during or after the study.

## Respiratory rate estimation

The normal range of resting RR for adults is 12-20 brpm[14]. Table 1 shows that the resting RRs of nine recruited participants were within the normal range, while the resting RRs of seven others were above normal.

The RR estimation algorithm successfully detected five out of the seven participants to be having higher than normal RRs. The other two, although their RRs were not identified as out of normal range, had been reported at the upper limit of the normal range (i.e., $\geq$ 19 brpm). The algorithm did not identify any of the actual normal range RRs as abnormal. Comparing the average one-minute RRs between the reference and RR estimation, we found that the smallest and largest mean absolute error of the RR algorithm is 0.2 brpm and 4.3 brpm, respectively. Autocorrelation function was introduced to capture the periodicity of the breath sounds as the dominant signals compared to the background noise which are normally less periodic in nature. Although this method works best for periodic signals, it has also been proven useful in extracting the periodicity of quasi-periodic physiological signals[15]. However, the accuracy of such method decreases with increasing inter-period variability of RR. The results showed that the RR algorithm achieved an overall high accuracy (with mean absolute error of $2.0\pm1.3$ brpm) among the 16 participants; the larger error that was reported could either be attributed to the difficulty in identifying an accurate reference for RR using a manual annotation process or the presence of high inter-variability of breath sounds.

The main challenge of the RR estimation algorithm is to accurately identify the inspiration and expiration of a breath and calculate the number of repeated patterns within a 15-second window. Only in cases where the breath patterns were consistent with distinctive inhale and exhale patterns (see the bigger and longer breath envelopes that were easily distinguished as exhales in Figure 1), the RR error was close to 0 brpm (i.e. 0.4 brpm).

*Table 1: Comparison of resting respiratory rate between the mask attachment and manual counting for spontaneous breathing.*

| Participant | Respiratory rate (brpm) | | Mean absolute error (brpm) |
|---|---|---|---|
| | Manual count | Mask Attachment | |
| 1 | 17.5 | 16.2 | 1.3 |
| 2 | 23.7 | 22.0 | 1.7 |
| 3 | 17.6 | 15.9 | 1.7 |
| 4 | 17.6 | 17.8 | 0.2 |
| 5 | 14.6 | 16.5 | 1.9 |
| 6 | 18.4 | 16.2 | 2.2 |
| 7 | 18.7 | 17.6 | 1.1 |
| 8 | 22.6 | 20.8 | 1.8 |
| 9 | 23.4 | 19.8 | 3.6 |
| 10 | 15.8 | 15.6 | 0.2 |
| 11 | 22.0 | 19.0 | 3.0 |
| 12 | 19.4 | 18.6 | 0.8 |
| 13 | 19.8 | 15.8 | 4.0 |
| 14 | 24.2 | 20.8 | 3.4 |
| 15 | 25.0 | 20.7 | 4.3 |
| 16 | 22.3 | 23.0 | 0.7 |
| | | **Average error** | **2.0 ± 1.3** |

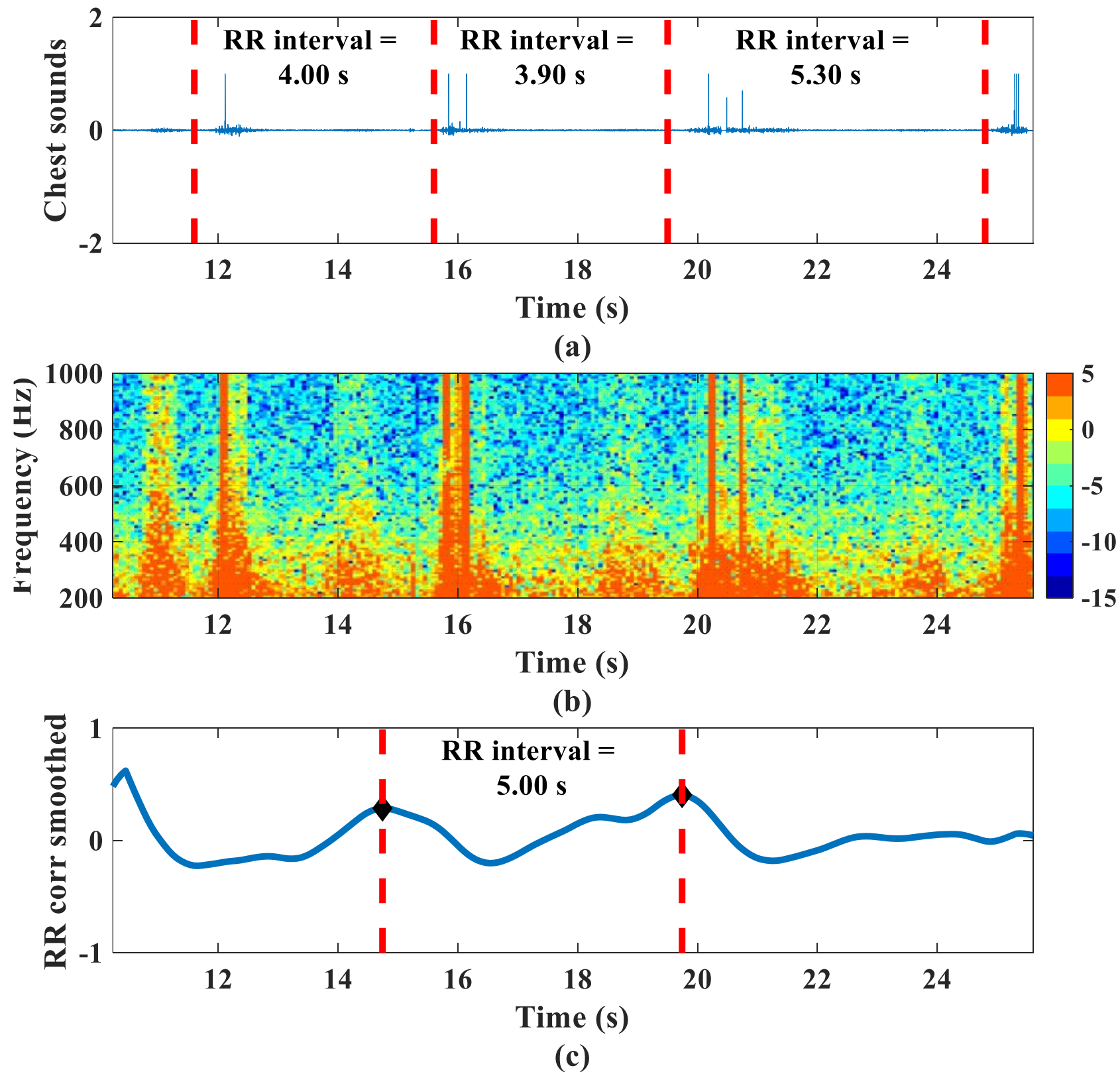

*Figure 1: An example of benchmarking the RR algorithm with the manual counting of breath cycles for a short segment of Participant 10's recording (ideal case). (a) Time domain plot, (b) time domain entropy plot of breath sounds, and (c) autocorrelation plot of entropy signals.*

Figure 2 demonstrates the risk of overestimating and underestimating the RR. In Figures 2(a) and 2(c), the actual RR was 12.9 brpm but the RR algorithm estimated as 18.4 brpm. Figure 2(b) shows that there was a burst split in the exhale between 10.5 s and 12.5 s within the same breath cycle. This may have contributed to the overestimation of the RR algorithm as it considered this short period of breath cycle in the evaluation window. Conversely, Figures 2(d) and 2(f) demonstrate the risk of underestimating the RR. The actual RR was 15.1 brpm, but the RR algorithm estimated 7.1 brpm. Such error occurs because the breath patterns were irregular (higher variability in the duration of inspiration and expiration and period of breath cycles within the evaluation window) and that was challenging for the algorithm to detect the global peaks in the autocorrelation plot in Figure 2(f).

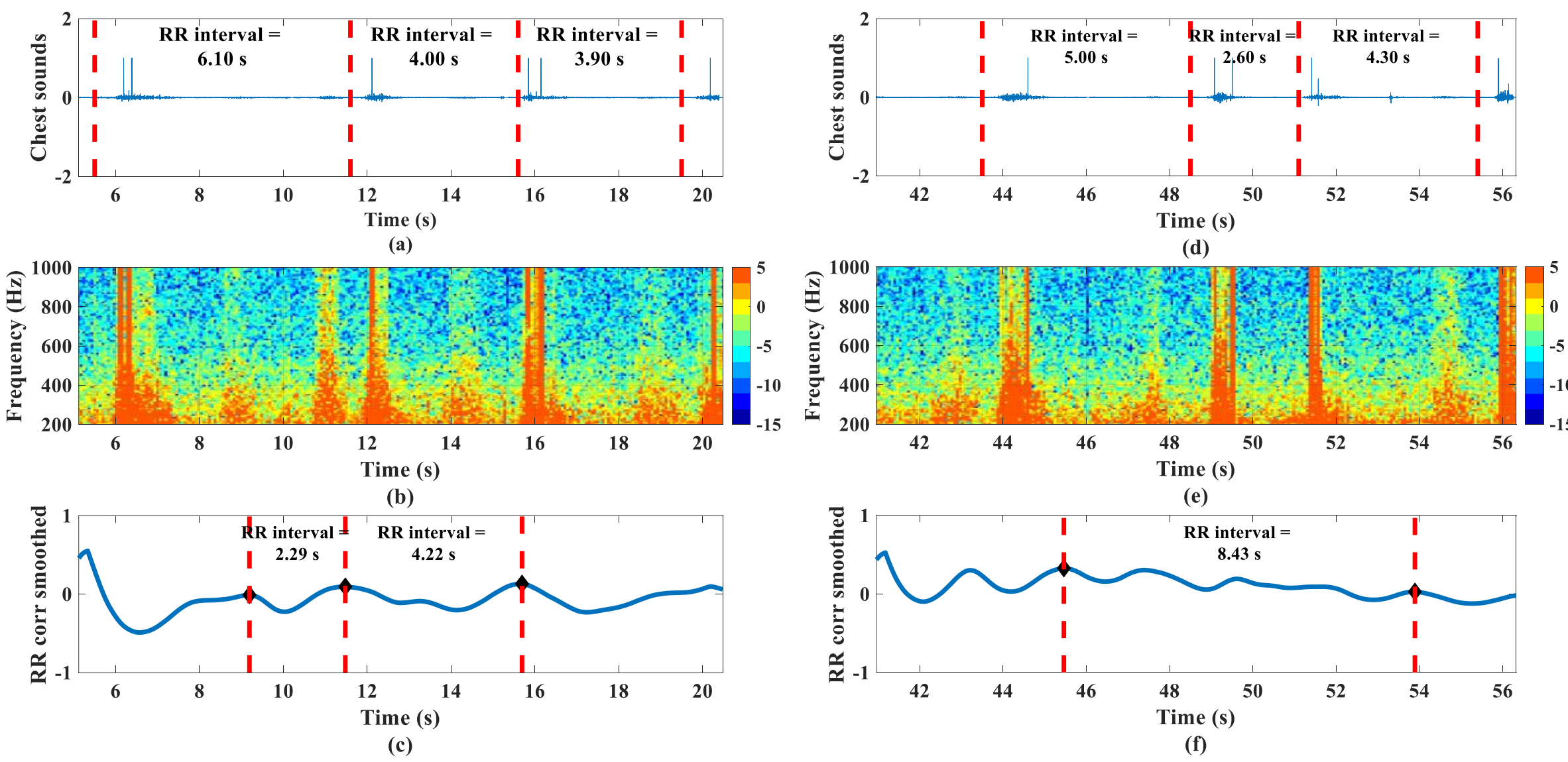

*Figure 2: A short segment plot of (a-c) participant 10's recording (estimated RR > actual RR) and (d-f) participant 10's recording (estimated RR < actual RR) in time domain between 200 Hz and 1000 Hz, time domain entropy plot of breath sounds, and autocorrelation plot of entropy signals.*

The gold standard of respiratory rate monitoring is capnometry, while the most common method used in the clinic is manual counting by healthcare professionals[16]. Even though there will be differences in accuracy between the outcomes of manual counting in 30-seconds and 60-seconds[17], our current annotation approach should account for the short-term respiratory rate variability to improve the reference's accuracy. However, manual annotations may also suffer from human error. A cross-sectional study has shown that even the manual assessments of RR by medical doctors were subjected to high inaccuracy when counting abnormally low and high RRs[18]. Manual counting of RR is also labor-intensive. Therefore, future larger sample size studies should validate the RR estimation algorithm with capnometry.

## Wheeze detection

The WZ detection algorithm was originally designed to work on chest sounds, using two low-complexity time-based spectral entropy features to differentiate wheezes from normal breaths. In the present study, we have added another feature, frequency-based entropy difference, to the WZ detection algorithm to improve classification accuracy. Figure 3 shows the distribution of normal breaths and wheezes in the feature space. The sensitivity, specificity, and accuracy of the simulated tracheal normal breaths and wheeze classified using a support vector machine (SVM) model with a radial basis kernel function were 88.8%, 94.9%, and 91.9%. respectively, as listed in Table 2. As the first introduction of this use case, the overall model performance is deemed acceptable although it can be further improved to increase the model fidelity as we collect more wheeze and normal breaths in the future. The present wheeze detection algorithm also achieved a balanced trade-off between sensitivity and specificity, which is another important criterion for a "good classifier". The model performance was evaluated using a 10-fold cross validation method to show that the model was robust when tested in 10 different scenarios.

Even though the use of a mannequin with a speaker does not represent an actual wheezy patient, it is a reasonable simulation that shows the direction of sound from the chest level through the airway to the sensor attached to the mask. In future studies, the wheeze detection algorithm should be validated in a clinical trial with actual wheezing patients.

The current study shows that the acoustic sensor is sufficiently versatile to be adapted for a different application, i.e., a mask attachment. Functionally, the original intended use of the sensor at the chest level provides a more comprehensive cardiopulmonary monitoring as it also estimates heart rate along with RR. However, the current proposed application provides an easier mode of administration and a more comfortable solution (with no direct surface contact of the sensor with the patient) to remotely monitor RR and detect wheeze. Notably, RR has also been recognized as the first indication as well as the best marker of deterioration in patient condition[19]. Therefore, the proposed mask attachment is helpful in managing patient conditions in overwhelmed clinics.

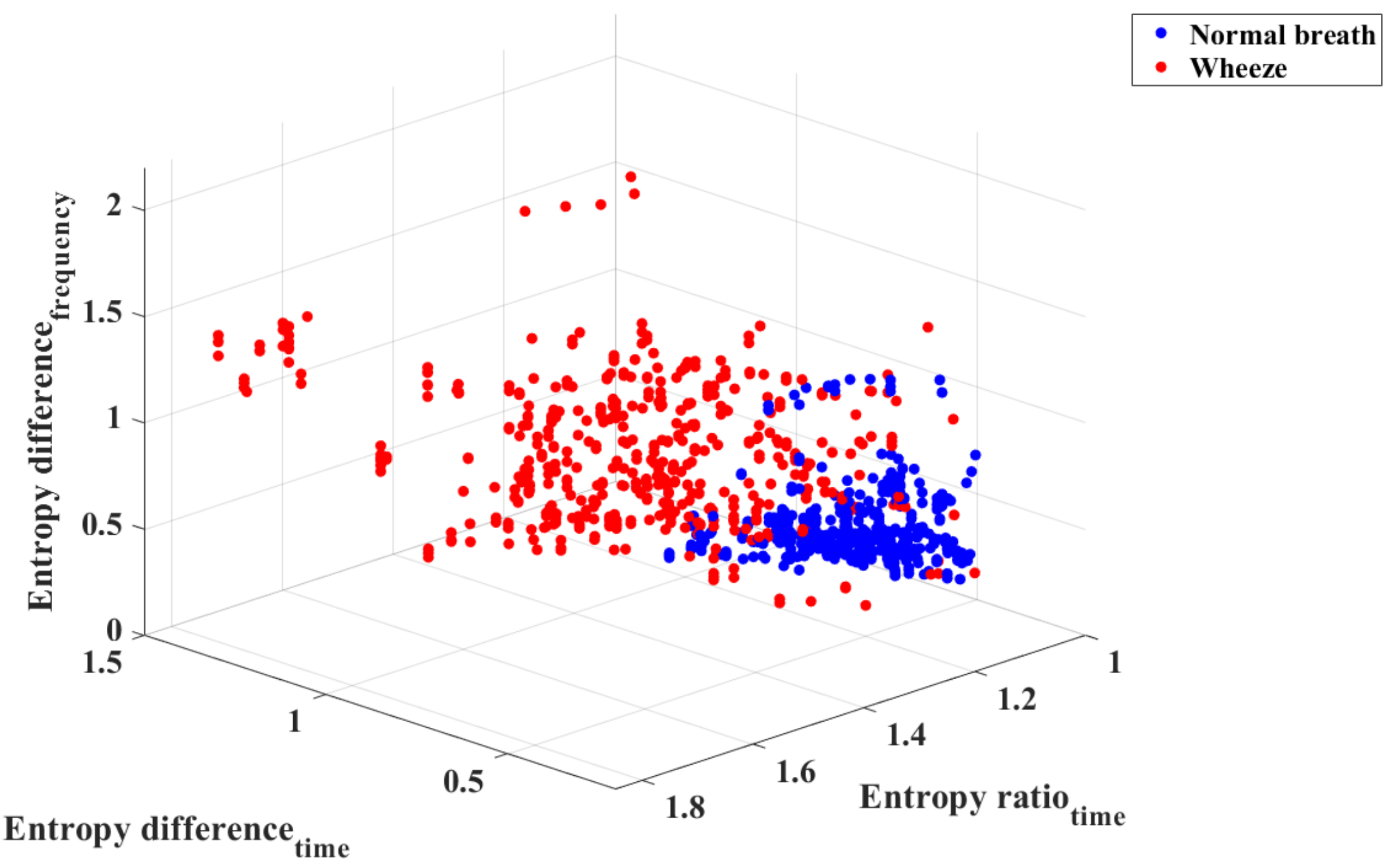

*Figure 3: Scatter plot of the three-dimension spectral entropy features of normal breath and wheeze.*

*Table 2: Performance of wheeze detection algorithm.*

| **Actual breath classification** | **Classification by wheeze detection algorithm** | | |
|---|---|---|---|
| | Wheeze | Normal breath | |
| **Wheeze** | 469 | 59 | Sensitivity: 88.8% |
| **Normal breath** | 27 | 501 | Specificity: 94.9% |
| | Positive predictive value: 94.6% | Negative predictive value: 89.5% | Accuracy: 91.9% |

# Methods

## Experimental setup

KN95 surgical respirators (also known as the face mask in the study) were modified by substituting the round plastic enclosed air valve with a mask attachment, as shown in Figure 4. A detailed description of the sensor is available in reference[13]. Sixteen participants were recruited for the study. Each participant's breath sound was recorded for one minute in an office room for the purpose of evaluating respiratory rate. The room was reasonably simulated as a clinic, subjected to a common workplace noise level. Participants were requested to breathe normally throughout the data collection.

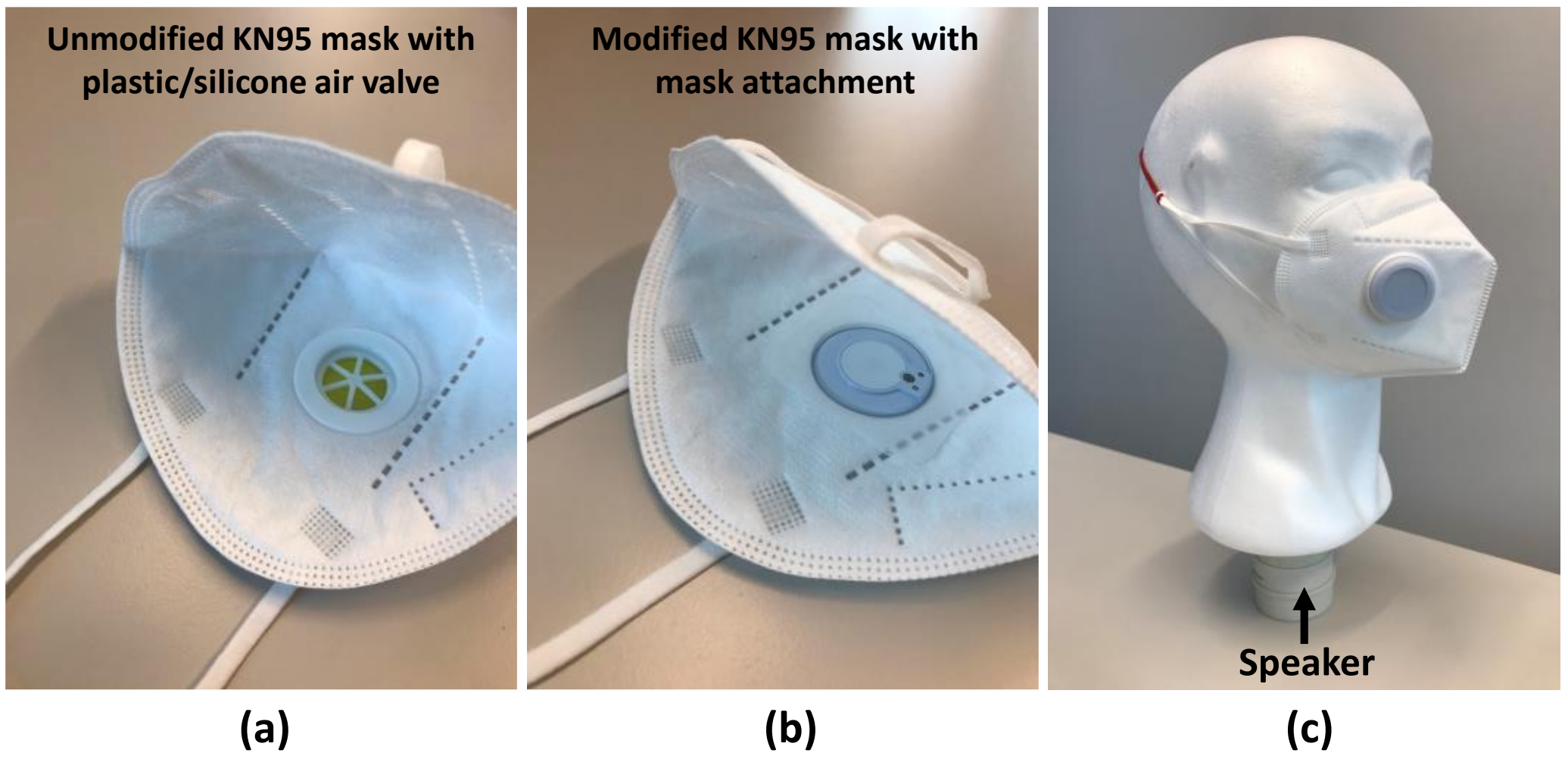

*Figure 4: Experimental setup for both the user test and simulated wheeze and normal breath signal through the mannequin for the wheeze detection test. (a) The unmodified KN95 mask with a plastic/silicone air valve. (b) The modified KN95 mask with a miniature sensor inserted through a silicon patch. (c) A speaker was inserted at the bottom of the mannequin to mimic the travel of lung sound through trachea to the face level.*

All participants underwent the study blindfolded, with the procedure described in Figure 5. None experienced any respiratory complications during data collection.

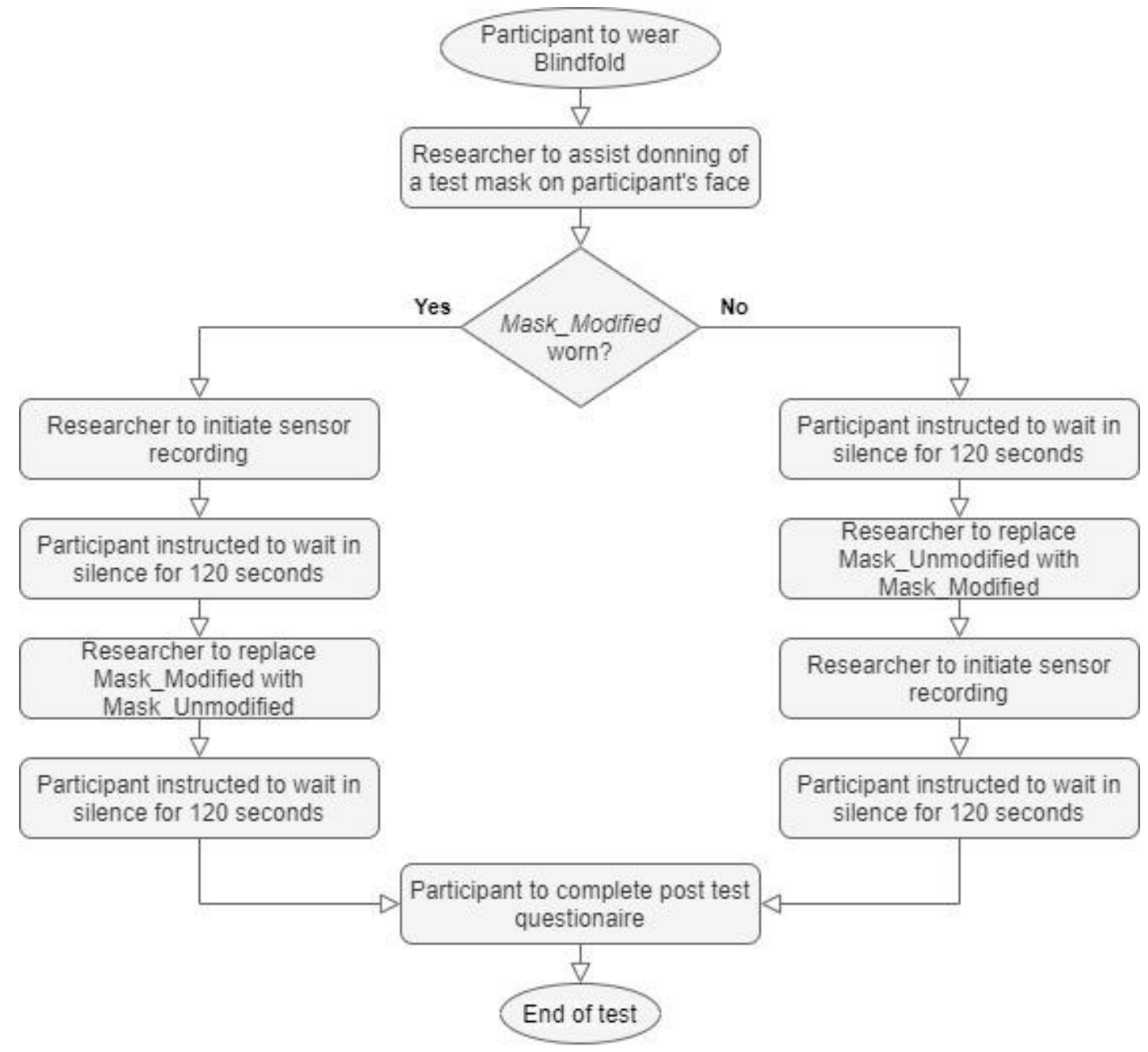

*Figure 5: Usability test procedure.*

## Human factor considerations

For the proposed method to be readily accepted by the intended users, implementation of the technology should not hinder the user's range of motion or affect normal activities. Hence, the mask attachment must be minimally intrusive to encourage compliance. Since the wearer of the attachment would likely already be required to be wearing a facemask, this segment of the study focuses on the resulting implication of wearing the attachment on top of a common facemask. The translation of a series of human factor considerations in three categories of use environment, user and use interface, into a desired outcome and iteration of the process is well described by the United States Food Drug Administration's Guidance document[20].

### *Use environment*

The mask attachment is proposed for the application of patient management in facilities where there is a large volume of patients. The environment can then be either in a controlled indoor or outdoor medical facility where a wireless internet connection is available. Temperature and humidity of the environment are directly related to the comfort of mask wear[21]. Also, the mask attachment should not significantly lower breathability or promote additional condensation formation within the mask. The inclusion of an additional attachment on a facemask should also not impede communication capability through speech. Therefore, the mask attachment must only alter a small area of the facemask.

### *User*

Medical grade facemasks, such as KN95 or N95 masks, come in various sizes and shapes to cater to the variant of possible face shapes and sizes. Sizing of these masks ensures proper fitment to achieve the intended outcome of air filtration for the wearer. Any attachment designed to be used with any mask should be physically small enough to be fitted onto variation of facemasks.

### *Use interface*

The comfort, safety, and ease of use of the mask attachment are the main concern when designing the use interface. The proposed attachment is 8 g in weight, 31 mm in diameter. and 10 mm in height. The modified mask weighs 14 g, while the unmodified mask weighs around 8 g unaltered and is generally accepted by wearers. With the addition of an attachment, the increase in weight of the mask can impact the comfort of wear. In addition, the physical characteristics of the attachment must not result in significant impedance of respiration for the wearer by obstruction of airflow.

## Participants

A sample of 16 adult participants was recruited into the study, with the participant profiles summarized in Table 3. All participants have been wearing a face mask regularly for at least 6 months due to the COVID-19 pandemic.

*Table 3: Participant profiles*

| **Attributes** | **Population n=16** |
|---|---|
| Age (average) | 24.5 ±4.0 |
| Gender | |
| Male | 5 |
| Female | 11 |
| Mask Size* | |
| S (< 100mm) | 1 |
| M (100mm – 110mm) | 11 |
| L (110mm – 125mm) | 4 |
| Pre-existing chronic respiratory conditions | 1 |

**Distance from middle of eyes to bottom of chin*

## Respiratory rate estimation and wheeze detection algorithms

In brief, the RR estimation algorithm is based on evaluating a 15-second breath segment's entropy autocorrelation. The collected breath sounds undergo a short-time Fourier-transform (STFT) by taking the fast Fourier transform (FFT) using a Hanning window size of 128 ms. The overlap size was 64 ms. The bandwidth for the entropy calculation was between 200 Hz and 1000 Hz. The entropy of each window frame was calculated using the principle of Shannon's entropy. More details of the RR estimation algorithms are available in reference[13].

Conversely, the WZ algorithm is based on evaluating a smaller breath segment - two and a half seconds. Furthermore, the window sizes of FFT and the Hanning window were comparatively smaller, at 32 ms. The overlap size was 90% of the FFT window size. The bandwidth for the entropy calculation was between 100 Hz and 2000 Hz. Three features were extracted from the entropy data: 1) time-based spectral entropy difference (maximum-minimum), 2) time-based spectral entropy ratio (maximum/minimum), and 3) frequency-based spectral entropy difference (maximum-minimum). The difference between the frequency-based and time-based spectral entropy features is that the entropy function was calculated in each of the row vector (frequency-based) and column vector (time-based) and the statistics (i.e., entropy difference and entropy ratio) were calculated across the row vectors (frequency-based) and column vectors (time-based) of the STFT outputs.

As no unhealthy participants who presented wheezes were recruited, a further simulation was employed to evaluate the WZ detection method. In the simulation, a speaker was used to play 176 random sample tracheal wheezes and 176 random sample normal breaths from an online database[22], through a foam mannequin, as shown in Figure 4. The same sound samples were played for three times to account for varying environmental noise, thus there were a total of 528 wheeze samples and 528 normal breath samples included in the data analysis. The modified KN95 mask with the sensor attachment was worn on the mannequin to collect wheezes and normal breaths.

## Validation of respiratory rate estimation and wheeze detection algorithms

The RRs calculated by the system were benchmarked with manual counts of breath periods within the 15-second evaluation windows. In the manual counting procedure, the start of each breath was annotated with a start time so that the duration of each breath cycle was derived as the interval between two consecutive start times. The annotations were done via cross-validations between visual and audio inspections. In doing so, the respiratory rate variability was accounted for in the manual counting method to make a comparable comparison with the RR estimation algorithm. To account for the robustness of the wheeze detection classification method toward inter-subject variability, 132 of the 176 wheezes and 132 of the 176 normal breaths, together with their respective two repeated measures were isolated for the acquisition of the SVM model. During this process, the data samples were subdivided into 70% of training samples and 30% of testing samples. The performance of the SVM model was cross-validated using a 10-fold cross validation method and the final model was acquired. To further validate the capability of the wheeze detection algorithm on new data, the remaining data that had not been used in the acquisition of the SVM model were included in the final evaluation of model performance. The sensitivity and specificity of the WZ detection algorithm were calculated from the percentage of true positive (wheeze) detected and the percentage of true negative (normal breath) detected, respectively. The dataset used for the WZ detection test was pre-annotated for the presence and absence of wheeze by the database[22].

## Exclusion criteria

In segments whereby the ground truths of RR were not obtainable, the results obtained from the mask attachment were not included in the analysis. The average of all the other segments where references were obtainable within the one-minute recording was computed and included in Table 1 for validation of the RR algorithm.

## Conclusion

Encountering the pandemic of COVID-19, donning a face mask becomes essential and even compulsory in certain countries as part of the effort in curtailing the spread of the virus among the community. We propose attaching an acoustic sensor on the face mask for the purpose of mass patient monitoring, complimenting hospital and clinic patient management efforts for large patient volume. We have evaluated the comfort of using the modified face mask through the verdict of 16 participants in the usability study. All participants reported that there was no discomfort from using the modified masks although about half of the participants did feel differences between the unmodified and modified masks. Lastly, the overall performances of RR estimation and WZ detection algorithms have shown that the sensor system, first developed for chest-worn sensing, adopted into this work as mask attachment is sufficiently versatile to be used as a mask attachment. Future validation of the application in a real clinical scenario with a larger sample size may be warranted to further increase the confidence, acceptability, and favor among the practitioners in clinics.

## Author contributions

VCA Koh, YY Ang, and RX Tan conceptualized, designed, and conducted the study. All authors analyzed and discussed the results. VCA Koh and YY Ang prepared the manuscript. RX Tan and W Ser reviewed and revised the manuscript.

## Competing interests

This research is funded by Aevice Health Pte Ltd, a medical device research and development company based in Singapore. Dr. Vivian CA Koh, Dr. Yi Yang Ang, and Dr. Rex X. Tan are employees of Aevice Health Pte Ltd. Dr. Rex X. Tan and Dr. Wee Ser are shareholders of Aevice Health Pte Ltd. A significant portion of the technology presented in this work is patent pending in Singapore with the application number 10202004626V filed by Aevice Health Pte Ltd.

## Data availability statement

The datasets generated during and/or analyzed during the current study are available from the corresponding author on reasonable request.

## Declaration statement

All methods were carried out in accordance with the Human Biomedical Research Act of Singapore 2015, revised in 2019. With reference to part 1, subsection 3.2 of the Human Biomedical Research Act of Singapore, part of this study where human participants was involved is not intended to investigate the prevention, prognostication, diagnosis or alleviation of any disease, disorder or injury affecting the human body; the restoration, maintenance, or promotion of the aesthetic appearance of human individuals through clinical procedures or techniques; or the performance or endurance of human individuals. Only healthy participants were recruited, and information collected are simple recordings of their breathe sound. Therefore, an approval of the protocol from an institutional board or licensing is not deemed to be required. All participants are employees of Aevice Health Pte Ltd, Singapore. An informed consent was obtained from all participants.